\documentstyle{aipproc}
\def\slash#1{\ooalign{\hfil/\hfil\crcr$#1$}}
\begin{document}
\title{Quark- and Gluon-Condensate Contributions to Penguin Four-Fermi Operators}

\author{Mohammad R. Ahmady\footnote{Email: mahmady2@julian.uwo.ca}, Victor Elias\footnote{Email:velias@julian.uwo.ca}}
\address{Department of Applied Mathematics, University of Western Ontario\\
London, Ontario, Canada N6A 5B7}

%\lefthead{LEFT head}
%\righthead{RIGHT head}
\maketitle

\begin{abstract}
The nonperturbative content of the QCD vacuum permits the occurrence of QCD-vaccum condensate contributions to penguin amplitudes.  We calculate the dimension-4 $<m_q\bar qq>$ and $<\alpha_s G^2>$ contributions to the effective Wilson coefficients for penguin four-Fermi operators, and discuss how such contributions may contribute to nonleptonic B decays. 
\end{abstract}

Nonleptonic B decays are very important to our understanding of standard model physics.  In particular, decays of the form $B\to h_1h_2$, where $h_1$ and $h_2$ are light hadrons $\pi$, $K$, $\eta$, $\eta' ,...$, necessarily entail an operator product expansion framework with field theoretical coefficients $C_i(\mu )$: 
\begin{equation}
<h_1h_2\vert H_{\rm eff}\vert B>=\frac{G_F}{\sqrt{2}}\sum_iV^i_{\rm CKM}C_i(\mu ) <h_1h_2\vert O_i(\mu )\vert B> \;\; .
\end{equation}
The standard model's heavy degrees of freedom associated with the top quark and the $W^\pm$ have been integrated out of the effective Hamiltonian appearing in (1).  The validity and utility of this description, as well as the effective truncation of higher dimensional operators from the expansion (1), must ultimately be tested empirically.  For the case $B\to h_1h_2$, it is standard to focus on the dimension-6 four-Fermi operators $O_{1-6}$, where operators $O_{1u,1c,2u,2c}$ are current-current operators and $O_{3-6}$ are penguin operators as given below, with $q$ being $\{ s,d\}$, respectively for $b\to \{ s,d\}$ transitions[$L\equiv 1-\gamma_5$, $R\equiv 1+\gamma_5$]:
\begin{eqnarray}
\nonumber O_{1u}&=&\bar q_\alpha\gamma^\mu Lu_\alpha\bar u_\beta\gamma_\mu Lb_\beta \; ,\; 
O_{2u}=\bar q_\alpha\gamma^\mu Lu_\beta\bar u_\beta\gamma_\mu Lb_\alpha \;\; ,\\
\nonumber O_{1c}&=&\bar q_\alpha\gamma^\mu Lc_\alpha\bar c_\beta\gamma_\mu Lb_\beta \; ,\; 
O_{2c}=\bar q_\alpha\gamma^\mu Lc_\beta\bar c_\beta\gamma_\mu Lb_\alpha \;\; ,\\
\nonumber O_3&=&\bar q_\alpha\gamma^\mu Lb_\alpha\sum_{q'}\bar q_\beta'\gamma_\mu Lq_\beta' 
\; ,\; O_4=\bar q_\alpha\gamma^\mu Lb_\beta\sum_{q'}\bar q_\beta'\gamma_\mu Lq_\alpha' \;\; ,
\\ O_5&=&\bar q_\alpha\gamma^\mu Lb_\alpha\sum_{q'}\bar q_\beta'\gamma_\mu Rq_\beta '\; ,\; 
O_6=\bar q_\alpha\gamma^\mu Lb_\beta\sum_{q'}\bar q_\beta'\gamma_\mu Rq_\alpha' \;\;.
\end{eqnarray}

Methodologically, there exists a genuine issue as to how to partition radiative corrections between coefficients and the matrix elements in (1).  We take the usual approach of allowing the matrix elements $<h_1h_2\vert O_i(\mu )\vert B>$ in (1) to be evaluated at tree order, with all higher order diagramatic perturbative corrections incorporated in Wilson coefficients $C_i(\mu )$.

We are particularly concerned with condensate contributions to penguin diagrams, in which an intermediate-state quark-antiquark pair emits an off-shell gluon that creates a final-state quark-antiquark pair.  Such contributions occur by virtue of the nonperturbative content of the QCD vacuum, which necessarily entails nonvanishing condensate-dependent expressions for QCD-vacuum expectation values of normal-ordered products of fields, quantities which would ordinarily vanish if the vacuum were purely perturbative\cite{pt}.  Thus the purely perturbative diagram (Fig. 1) must be augmented by diagrams sensitive to the nonperturbative vacuum expectation values of quark-antiquark pairs (Fig. 2) and multiple gluon fields (Fig. 3).

The purely perturbative Fig. 1 contributions to penguin operators are already known to next-to-leading (NLL) precision via naive dimensional regularization within an $\overline{\rm MS}$ context \cite{kps}.  We choose to incorporate such corrections entirely within the Wilson coefficients $C_{3-6}$, which are now denoted to be "effective" Wilson coefficients as follows:
\begin{eqnarray}
C_3\rightarrow C_3^{\rm eff}&=&C_3-\frac{1}{6}\Delta C\;\; , \\
C_4\rightarrow C_4^{\rm eff}&=&C_4+\frac{1}{2}\Delta C\;\; , \\
C_5\rightarrow C_5^{\rm eff}&=&C_5-\frac{1}{6}\Delta C\;\; , \\
C_6\rightarrow C_6^{\rm eff}&=&C_6+\frac{1}{2}\Delta C\;\; ,
\end{eqnarray}
with
\begin{eqnarray}
\nonumber
\Delta C&=& \frac{\alpha_s}{4\pi}\left\{ \left [\frac{4}{3}+\frac{2}{3}Ln(\frac{m^2_{s}}{\mu^2})+\frac{2}{3}Ln(\frac{m^2_{b}}{\mu^2})-\Delta F_1(\frac{k^2}{m^2_{s}})-\Delta F_1(\frac{k^2}{m^2_{b}})\right ]C_3 \right .\\
\nonumber
&{}&+\sum_{q'=u,d,s,c,b}\left [\frac{2}{3}Ln(\frac{m^2_{q'}}{\mu^2})-\Delta F_1(\frac{k^2}{m^2_{q'}})\right ](C_4+C_6) \\
&{}&\left . -\sum_{q'=u,c}\frac{V_{q'b}V_{q'q}^*}{V_{tb}V_{tq}^*}\left [\frac{2}{3}+\frac{2}{3}Ln(\frac{m^2_{q'}}{\mu^2})-\Delta F_1(\frac{k^2}{m^2_{q'}})\right ]C_1 \right \}\;\; , 
\end{eqnarray}
\begin{equation}
\Delta F_1(z)\equiv -4\int^1_0dxx(1-x)ln\left [1-zx(1-x)\right ]\;\; .
\end{equation}
An important feature of such purely perturbative corrections to penguin operators is their dependence not only on the internal quark mass $m_{q'}$, but also on the value of $k^2$ characterizing the final-state quark-antiquark pair.  This behaviour is particularly important for CP violation, since $\Delta F(k^2/m^2_{q'})$ acquires an imaginary part when $k^2>4m^2_{q'}$.  Such an imaginary part has already been suggested as the origin of the strong phase necessary for the generation of direct CP asymmetry\cite{kps}.

The occurrence of $k^2$-dependent effective Wilson coefficients is also a feature of nonperturbative corrections arising from QCD vacuum condensates.  Such condensate effects, which are already understood via their QCD sum-rule duality to the hadron spectrum, enter radiative corrections to penguin-mediated processes via Figs 2 and 3.  To obtain the quark-condensate contribution to such processes (Fig 2), we proceed by replacing the usual 
perturbative internal quark propagator $(S^{\rm P})$ in Fig. 1 with the full quark propagator 
$S(p)$:
\begin{equation}
S(p)=S^{\rm P}(p)+S^{\rm NP}(p).
\end{equation}
The nonperturbative contribution to the quark propagator $(S^{\rm NP})$ is just
\begin{equation}
iS^{\rm NP}(p)=\int d^4xe^{ip.x}<\Omega\vert :\Psi (x)\bar\Psi (0):\vert\Omega >\;\; ,
\end{equation}
where $\vert\Omega >$ is the nonperturbative QCD vacuum\cite{baes}. 
The nonlocal vacuum expectation value (vev) in eq. (10) can be expanded in terms of 
local condensates.  We note that the nonlocal vev of two quark fields does not contain a gluon condensate component\cite{ess}.  The quark condensate projection of the nonperturbative quark propagator is taken from Ref. \cite{baes}:
\begin{equation}
iS^{\rm NP}(p)={(2\pi )}^4(\slash{p} +m_q)F(p)\;\; ,
\end{equation}
where the Fourier transform of $F(p)$ is expressed in terms of Bessel function
\begin{equation}
\int d^4pe^{-ip.x}F(p)=-\frac{<\bar qq>}{6m_q^2}\frac{J_1(m_q\sqrt{x^2})}{\sqrt{x^2}}\;\; ,
\end{equation}
with the following mass-shell property:
\begin{equation}
(p^2-m_q^2)F(p)=0\;\; .
\end{equation}
The dimension-3 quark condensate $<\bar qq>$ is defined to be the vacuum expectation value of the normal ordered local two-quark fields, i.e.
\begin{equation}
<\bar qq>=<\Omega\vert :\bar\Psi (0)\Psi (0):\vert\Omega >\;\; .
\end{equation}
Using the Feynman rule of the eq. (11), one can obtain the nonperturbative \mbox{$<\bar qq>$} contribution to the effective Wilson coefficients in a straightforward manner.  The relevant Feynman diagram is illustrated in Fig. 2 where the nonperturbative quark propagator $S^{\rm NP}$ is depicted by a disconnected line with two dots.   Here we concentrate on the loop portion of Fig. 2 which differs from the purterbative case of Fig. 1.  Aside from the color factor, the vector current correlation function can be written as
\begin{equation}
\Pi_{\mu\nu}^{<\bar q q>}(k)=2\int d^4p\frac{Tr[(\slash{p}-\slash{k}+m)\gamma_\mu (\slash{p}+m)\gamma_\nu ]F(p)}{{(p-k)}^2-m^2+i\epsilon}\;\; ,
\end{equation}
where the factor 2 in front is due to two possible insertions of the nonperturbative quark propagator in the fermion loop.  By contracting $p^\mu p^\nu$ into $\Pi_{\mu\nu}^{<\bar qq>}$ and using the identity\cite{baes}
\begin{eqnarray}
\nonumber
\int d^4pk.pF(p)&=&i\lim_{\xi\to 0}\frac{d}{d\xi}\int d^4pe^{-i\xi k.p} F(p)\\
&=&-i\frac{<\bar qq>}{6m_q}\lim_{\xi\to 0}\frac{d}{d\xi}\frac{J_1(m_q\xi\sqrt{k^2})}{m_q\xi\sqrt{k^2}}=0\;\; ,
\end{eqnarray}
where the second line is derived from eq. (12), one can show explicitly the transversality of the correlation function in eq. (15).  Consequently, one finds upon contracting $g^{\mu\nu}$ into $\Pi_{\mu\nu}^{<\bar qq>}$ and imposing the on-shell constraint (13) that
\begin{eqnarray}
{\Pi^{<\bar qq>}}_\mu^\mu (k^2) =24(2m_q^2+k^2)\int d^4p\frac{F(p)}{k^2-2k.p+i\epsilon}-24\int d^4pF(p)\;\; .
\end{eqnarray}
The integrals appearing in eq. (17) are evaluated as follows\cite{baes}:
\begin{eqnarray}
\nonumber
\int d^4pF(p)&=&-\frac{<\bar qq>}{12m_q}\;\; , \\
\nonumber
\int d^4p\frac{F(p)}{k^2-2k.p+i\epsilon}&=&\frac{i<\bar qq>}{12m_q^2\sqrt{k^2}}\int^\infty_0\frac{d\eta}{\eta}e^{i\eta k^2-\epsilon\eta}J_1(2\eta m_q\sqrt{k^2}) \\
&=&-\frac{<\bar qq>}{6m_qk^2}{\left [1+\sqrt{1-\frac{4m_q^2}{k^2}}\right ]}^{-1}\;\; ,
\end{eqnarray}
where the final line is derived by utilizing a tabulated integral\cite{gr}.  Using the transversality of $\Pi_{\mu\nu}^{<\bar qq>}$, the result for the RHS of (15) is easily obtained:
\begin{equation}
\Pi_{\mu\nu}^{<\bar qq>}=-\frac{<\bar qq>}{3m_q^3}\left [1-\left (1+\frac{2m_q^2}{k^2}\right )\sqrt{1-\frac{4m_q^2}{k^2}}\right ](g_{\mu\nu}k^2-k_\mu k_\nu )\;\; .
\end{equation}
This result is, in fact the $<\bar qq>$-contribution to the vector-current correlation function first derived in ref. \cite{blp}.  The corresponding gluon condensate contribution to the vector-current correlator, which may be derived by several different methods \cite{baes,blp}, is given by
\begin{eqnarray}
\nonumber
\Pi^{<G^2>}_{\mu\nu}&=&-\frac{\alpha_s <G^2>}{48\pi m_q^4}(g_{\mu\nu}k^2-k_\mu k_\nu )\frac{{(\frac{m_q^2}{k^2})}^2}{{(1-\frac{4m_q^2}{k^2})}^2} \\
& &\times  \left [\frac{48{(\frac{m_q^2}{k^2})}^2(1-\frac{2m_q^2}{k^2})}{\sqrt{1-\frac{4m_q^2}{k^2}}}Ln\frac{\sqrt{1-\frac{4m_q^2}{k^2}}+1}{\sqrt{1-\frac{4m_q^2}{k^2}}-1}-4+\frac{16m_q^2}{k^2}-48{(\frac{m_q^2}{k^2})}^2\right ]\;\; .
\end{eqnarray}
where
\begin{equation}
z \equiv k^2/m_q^2
\end{equation}
and where \cite{blp,asd}
\begin{eqnarray}
X(z) & \equiv & \frac{1}{(1-4/z)} \left[ \int_0^1 dx \;  ln [1 - z x (1-x) - i | \epsilon | ] + 2 \right] \nonumber\\
& = & \left[
\begin{array}{ll} \frac{1}{\sqrt{1+4/|z|}} ln \left[ \frac{\sqrt{1+4/|z|}\;+1}{\sqrt{1+4/|z|}\; -1} \right], & z < 0 \\
\frac{2\sqrt{4/z-1}}{(1-4/z)} tan^{-1} \left[ \frac{1}{\sqrt{4/z-1}} \right], & 0 < z < 4 \\ 
\frac{1}{\sqrt{1-4/z}} \left( ln \left[ \frac{1 + \sqrt{1-4/z}}{1-\sqrt{1-4/z}} \right] - i \pi \right), & z > 4
\end{array}
\right]
\end{eqnarray}
This contribution diverges sharply at $z=4$, a divergence which is {\it not} removable by the operator realignment suggested in ref. \cite{jm}.

We see that both $<\bar qq>$ and $<G^2>$ contributions to penguin operators occur via their known contributions to the vacuum polarization functions [eqs. (19) and (20)] within Figs 2 and 3. {\it in exactly the same way} that the purely perturbative vacuum polarization within Fig. 1 also enters penguin operators.  This latter contribution is responsible for the vacuum-polarization Feynman loop integral $\Delta F_1$, as defined by (8), which appears explicitly in the penguin-operator correction $\Delta C$ [eq. (7)].  Consequently, the aggregate effect of these lowest-dimensional condensate contributions to penguin operators is obtained by retaining equations (3-7), but with $\Delta F_1$ redefined so as to include $<\bar qq>$ and $<G^2>$ contributions to the vacuum polarization:
\begin{eqnarray}
\nonumber
\Delta F_1 (z)&=&-4\int^1_0dxx(1-x)Ln\left [1-zx(1-x)\right ] \\
&+& \frac{8\pi^2<m_q\bar qq>}{3m_q^4}\left [1-\left (1+\frac{2}{z}\right )\sqrt{1-\frac{4}{z}}\right ] \\
\nonumber &+& \frac{\pi <\alpha_s G^2>}{6m_q^4}\frac{1}{z^2{(1-\frac{4}{z})}^2}\left [ \frac{48(1-\frac{2}{z})}{z^2\sqrt{1-\frac{4}{z}}}Ln\frac{\sqrt{1-\frac{4}{z}}+1}{\sqrt{1-\frac{4}{z}}-1}-4+\frac{16}{z}-\frac{48}{z^2}\right ]
\;\; .
\end{eqnarray} 
Eq. (23) is expressed in terms of the renormalization-group invariant condensates (to one-loop order in $\alpha_s$) whose magnitude are known from QCD sum-rule applications.   For example, if $z$ is much larger than one, as would be the case for very light quarks [eq. (21)], the nonperturbative contributions to (23) are approximately given by
\begin{equation}
\Delta F_1^{\rm nonperturbative}\approx\frac{16\pi^2<m_q\bar qq>}{k^4}-\frac{2\pi <\alpha_s G^2>}{3k^4} \;\; .
\end{equation}
Using phenomenological values $<m_q\bar qq>=-f^2_\pi m_\pi^2/4$ \cite{gor} and $<\alpha_sG^2>=0.045$ GeV$^4$ \cite{svz}, these nonperturbative contributions are seen to be three orders of magnitude smaller than corresponding purely perturbative contributions to $\Delta F_1$.

Indeed, for $B\to h_1h_2$ decays, we may anticipate that $k^2$ is typically between $m^2_b/4$ and $m_b^2/2$ \cite{sw}.  This range, however, suggests sensitivity to the $z=4$ singularity mentioned above when the intermediate-state quark $q'$ in (7) is identified with the charmed quark [recall that $z$ in (23) can be identified with $k^2/m_{q'}^2$ in (7)].  One cannot, therefore, discount the possibility that the gluon-condensate contribution to the penguin amplitude may be comparable to the purely perturbative contribution.

This result has possible ramifications for investigations of CP asymmetry in hadronic B decays.  The gluon condensate contribution to (7)[via (23)] develops an imaginary part as $k^2\to 4m_c^2$ from above that would dominate the strong phase of the effective Wilson coefficients (3-6).  Interestingly, the $<m_q\bar qq>$ contribution to (23), though small, has an imaginary part over the range $0<z<4$, suggestive of an additional source of CP violation stemming from light quark-antiquark-pair intermediate states.  This latter absorptive contribution has been discussed in other contexts as a manifestation of the Goldstone theorem \cite{emss}.  The singularity of the gluon-condensate contribution to penguin operator coefficients at $z=4$, however, suggests a resonance interpretation.

Consider, for example, the dileptonic B decays $B\to X_s\ell^+\ell^-$.  Such decays acquire a continuum charmed quark loop contribution, analogous to the quark-loop in Fig. 1, in addition to a resonance contribution associated with the $c\bar c$ resonances $J/\psi ,\; \psi' ,\; ...$ \cite{ahm96}.  This resonance contribution can be modeled by the presence of a resonance propagator $1/(M^2-k^2-iM\Gamma )$ in the long distance amplitude, $M$ and $k$ being respectively the resonance-mass and momentum-transfer to the the lepton pair.  The presence of the resonance propagator leads to a peak at $k^2=M^2$ dominating the branching ratio for this decay mode.  For the $B\to h_1h_2$ case, the $z=4$ gluon-condensate singularity of $\Delta F_1$ within $\Delta C$ [eq. (7)] suggests the occurrence of a "charmonium" resonance at $k^2=4m_c^2$.  However, the fact that this intermediate state couples to a {\it gluon}, as opposed to the electromagnetic-penguin photon appropriate for $B\to X_s\ell^+\ell^-$, necessarily implies that this $c\bar c$ resonance be a weakly bound {\it colour octet} state.  We note that the idea of colour-nonsinglet intermediate states within QCD is quite standard; such states are, of course, quarks and gluons, which are prohibited from manifesting themselves as physical final states by additional dynamics (confinement).  What is new here is the idea of a (similarly confined) colour-nonsinglet bound state also contributing to a nonleptonic process via QCD's nonperturbative content, the QCD-vacuum condensate.

We reiterate, however, that the $q\bar q$ momentum-transfer $k^2$ is {\it not} an observable in the $B\to h_1h_2$ decay process, in contrast to the observable dilepton momentum-transfer in $B\to X_s\ell^+\ell^-$.  It should be noted, though, that $4m_c^2$ is not an unreasonable choice for the value of $k^2$ characterizing $B\to h_1h_2$, and that the gluon-condensate component of the decay amplitude may well play an important role both in enhancing such decays and in providing a mechanism for CP violation. 
\vskip 1.5cm
\noindent
{\bf\large Acknowledgement}\\
We are grateful for support from the Natural Sciences and Engineering research Council of Canada.

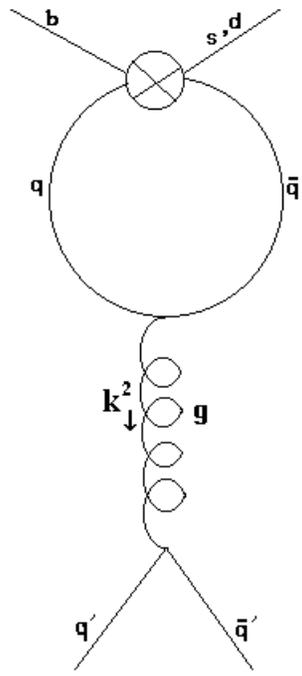
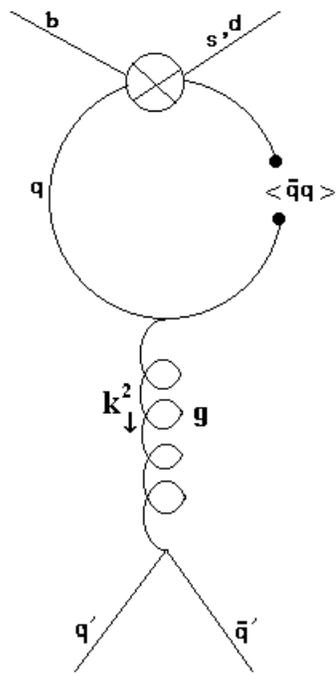
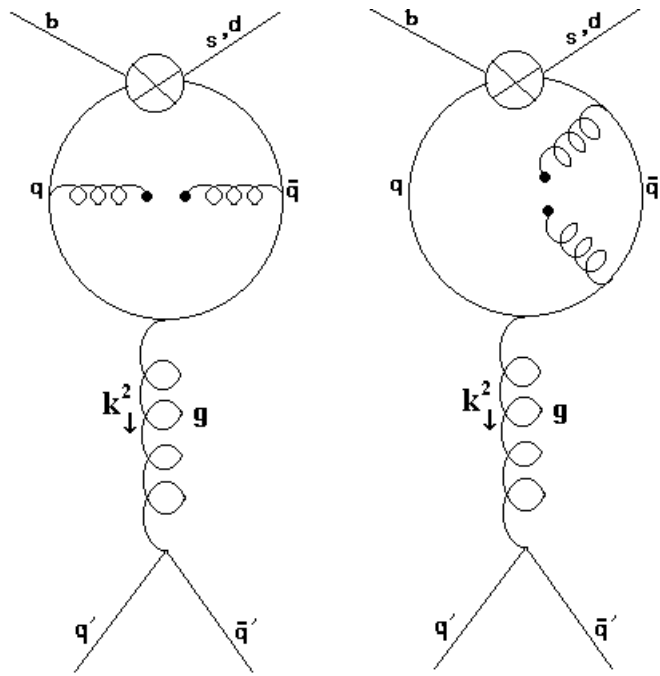
\end{document}